\newif\ifdraft\draftfalse
\newif\iffull\fullfalse

\drafttrue \fulltrue

\iffull
\documentclass[[a4paper,11point,envcountsame]{article}

\else
\documentclass[11pt]{article}

\fi
\usepackage{epsfig,url}
\usepackage{amsmath}

\usepackage{graphics}
\usepackage{color}
\usepackage{float}
\newcommand{\old}[1]{}                          

\begin{document}

\title{Timed Analysis of Security Protocols}

\author{R. Corin\footnote{Contact author's email is {\tt  ricardo.corin@utwente.nl}.
}, S. Etalle, P.H. Hartel and A. Mader\\
University of Twente, The Netherlands\\
}

\date{}

\maketitle

\begin{abstract}
  We propose a method for engineering security protocols that are
  aware of timing aspects.
  We study a simplified version of the well-known Needham Schroeder
  protocol and the complete Yahalom protocol, where timing information
  allows the study of different attack scenarios.  We model check the
  protocols using UPPAAL.  Further, a taxonomy is obtained by studying
  and categorising protocols from the well known Clark
  Jacob library and the Security Protocol Open Repository (SPORE)
  library.  Finally, we present some new challenges and threats that
  arise when considering time in the analysis, by providing a novel
  protocol that uses time challenges and exposing a timing attack over
  an implementation of an existing security protocol.

\end{abstract}



\section{Introduction}

The communication via a shared medium, like the Internet, is
inherently insecure: anyone has access to en route messages and can
potentially eavesdrop or even manipulate the ongoing
communication. Security protocols are distributed programs
specifically designed to achieve secure communication over such media,
typically exchanging messages built constructed using
cryptographic operations (e.g.~message encryption). 

Security protocols are difficult
to design correctly, hence their analysis is critical.  A
particularly successful model to analyze security protocols is the
Dolev~Yao model~\cite{DY83}, in which the attacker is assumed to have complete
control over the network. Also, the model assumes ideal
cryptography, where cryptographic operations are assumed to be
perfect.  The Dolev~Yao model is attractive because it can be easily
formalized using languages and tools based on formal
methods. Moreover, the model has an appropriate level of abstraction,
as many\old{security protocol} attacks\old{(and the attacks that
  exploit those errors)} are\old{completely} independent of the
underlying details of the cryptographic operations
and 
are based only on combinations of message exchanges plus knowledge
gathered by the attacker during the execution.

Typically, Dolev~Yao methods for formal verification of security protocols
(among the proposals~\cite{Lowe97,DE02,CE02}) do not take
\emph{time} into account, and this choice simplifies the analysis.
However, security protocols, like distributed programs in general,
are sensitive to the passage of time. Recently, consideration of time
in the analysis of security protocols has received some attention (see
Related Work below), but this attention has been focused mostly on
timestamps.

In this paper\footnote{An earlier version appeared in~\cite{Cor04}.}
 we develop an analysis model for security protocols that
explicitly takes into account \emph{time} flowing during the execution
of a protocol.  
%
%
In general, in the design and implementation of a security protocol
two aspects of timing must be considered at some stage:
\begin{enumerate}
\item 
Time can influence the flow of messages. For instance, when a message
does not arrive in a timely fashion (i.e. \emph{timeouts}),
retransmissions or other actions have to be considered.
\item 
Time information can be included within protocol messages (e.g. timestamps).
\end{enumerate}

Consider first (1) above.  In general, the influence of time on the
flow of messages is not usually considered by current state of the
art methods for analysing protocols. However, we believe it to be
crucial because (i) Even if the abstract protocol does not decide what
action to take at a particular moment of the execution (e.g. in the
case of timeouts), the actual implementation will eventually have to
consider these issues anyway; (ii) The efficiency and security of the
implementation depends critically on these specific decisions; and
(iii) The timing of message flows in a protocol can be exploited by an
attacker.  
\old{(As an example of such an exploit we describe a timing
  attack on an implementation of Abadi's Private Authentication
  protocol.)  }

Now consider item (2) above. There, we believe that making judicious
use of timing information in a protocol has received attention but
mostly in the limited setting of using time stamps as opposed to
nonces. However, time information can be used to influence message
flows as well, as we illustrate in Section~\ref{issues}.

\paragraph{Contributions}
Our study covers several issues in the study of time in security protocols. 

\begin{itemize}
\item
Firstly, in Section \ref{sec:time-retr} we study which kinds of timing
issues, like timeouts and retransmissions, may arise in the study of
security protocols. We then proceed in Section \ref{modeling} to
present a method for the design and analysis of security protocols
that consider these timing issues.  The method is based on modelling
security protocols using timed automata~\cite{AD94}.  In support of
the method we use UPPAAL~\cite{Amn00} as a tool to simulate, debug and
verify security protocols against classical safety goals like secrecy
and authentication, \emph{in a real time scenario}, using reachability
properties.  As examples, we analyse a simplified version of the
Needham Schroeder protocol~\cite{Lowe95} and the full Yahalom
protocol~\cite{CJ96} in Section~\ref{sec:analysing-protocols}.

\old{
Our developed analysis model is illustrated in
Figure~\ref{figtime}. The general analysis model of the Introduction
(shown in Figure~\ref{sm}) is here instantiated using timed automata,
which enables timed analysis.  In Figure~\ref{figtime}, the attacker
submodel implements a timed Dolev~Yao attacker, which can synchronize
on communications with the other participants' automata, as detailed
in Section~\ref{sec:modelling-adversary}. Also, the protocol
participants are represented as sequential timed automata. Then, both
the attacker and participants are composed in parallel to conform a
system that can be analysed to establish reachability
properties. These properties state either secrecy (i.e. whether it is
possible to reach a state in which the attacker automaton knows a
secret) or authentication (i.e. whether it is possible to reach a
state in which a participant automaton ends its execution believing
wrongly it communicates to some participant).
}
\item 

  Secondly, in Section~\ref{sec:taxon-mess-flows}, we categorize all
  the protocols from the Clark and Jacob library and the SPORE library
  into different (more abstract) patterns of message flows with
  timeouts. We then analyse each abstract pattern, independently of
  the actual protocols, and establish their timing efficiency and
  security.

\item
Finally, in Section \ref{issues} we illustrate some novel opportunities and
difficulties that appear when considering time in the design and
analysis of security protocols:

\begin{itemize}
\item
In Section~\ref{timedchal} we
give an example protocol that accomplishes authentication by
exploiting the \emph{timeliness} of messages.  The protocol uses time
in a conceptually new way, by employing \emph{time challenges} as a
replacement for nonces.  

\item 
As a second example of a novel difficulty in
Section~\ref{section:private}, we describe how timing
attacks~\cite{Eng96} can be applied to security protocols, by
describing an attack over Abadi's private authentication protocol
\cite{Aba00}. Although these protocols can be modelled as timed
automata, thus permitting general verification, we leave the detailed
verification as future work since for this we need a model checker
that is also \emph{ probabilistic} (like \cite{Arg01} or
\cite{KNP02b}): our nondeterministic intruder of UPPAAL is too
powerful, since it can always guess correctly times and values even if
the probability of guessing is negligible.

\end{itemize}
\end{itemize}

\section{Timeouts and Retransmissions}
\label{sec:time-retr}

To illustrate how time influences the analysis of security protocols
(even when it does not explicitly use timing information), consider
the following protocol written in the usual notation.
\begin{eqnarray*}
1. \ A \rightarrow B &:& M_{AB}\\
2. \ B \rightarrow A &:& M_{BA}
\end{eqnarray*}

\begin{figure}
\begin{center}
\input{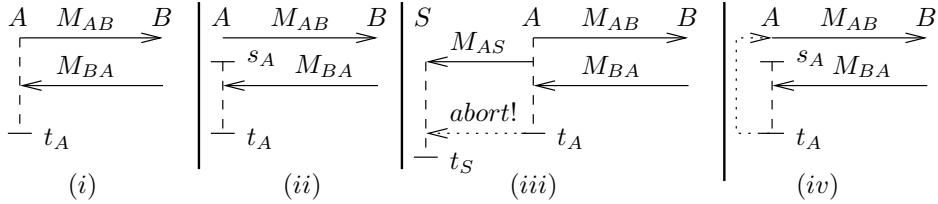}
\caption{\emph{Left}: Timeouts (i) typical (ii) windowed; \emph{Right}: timeout actions (iii) chained abort and (iv) retransmission }
\label{messageflows}
 \end{center}

\end{figure}

Here, first $A$ sends message $M_{AB}$ to $B$, and later $B$ sends
message $M_{BA}$ to $A$.  This high-level view does not consider
timing.  To consider time, we first need to assume that both $A$ and
$B$ have \emph{timers}. In this paper, we do not require timers
between parties to be synchronised (see below for a discussion).  The
next step consists in distinguishing the different operations that
occur, with their respective times. In Step 1, it takes some time
to create $M_{AB}$.  The other operation that takes time is the actual
sending of the message, ie. the time it takes $M_{AB}$ to travel
from $A$ to $B$. This transmission
time is unbounded, since the message may be lost or intercepted, and
therefore $A$ may need to \emph{timeout}: After $A$ sends $M_{AB}$,
she starts a timer that will timeout if $M_{BA}$ (Step 2 of the
above protocol) is not received after some waiting, say $t_A$ (Figure
\ref{messageflows} $(i)$).  Clearly, $t_A$ should be greater than the
time of creating $M_{BA}$, plus the average time of sending both
$M_{AB}$ and $M_{BA}$.  In general, $A$ does not need to start waiting
for a response immediately after sending a message; for instance, $A$
could hibernate (or start doing another task) for some time $s_A$
before beginning to expect the response $M_{BA}$. This results in a
\emph{windowed} timeout (Figure \ref{messageflows} $(ii)$).
Typically, the values for $s_A$ and $t_A$ depend on implementation
details.  However, an implementation independent quantitative analysis
could already give an early indication of what attacks can be mounted
for some values that are no longer possible for others (eg.  a smaller
$t_A$ and a larger $s_A$).

Another issue that is not considered either in previous approaches is
that the \emph{action} to be taken when a timeout occurs is sensitive.
Typically, the implicit assumption is that the protocol should abort,
as it is the case in Figure
\ref{messageflows} $(i)$.  This means that the protocol party that
reaches the timeout deduces that a fault has happened. 
However, aborting may not consist only of stopping execution
altogether.  For example, if we consider protocols with several
parties, we may wish that when a party timeouts it also communicates its
decision to abort to other, still active parties. For instance,
consider the following protocol:
\begin{eqnarray*}
1. \ A \rightarrow B &:& M_{AB}\ \ \ \ \ \text{  $A$ starts timer expecting $M_{BA}$}\\
2. \ A \rightarrow S &:& M_{AS}\ \ \ \ \ \text{  $S$ starts session timer} \\
3. \ B \rightarrow A &:& M_{BA}
\end{eqnarray*}

Here, if $A$ times out on Step 2, she could communicate the abort
decision to $S$, as shown in in Figure \ref{messageflows} $(iii)$.

Aborting execution is not the only feasible action to perform after a
timeout~\cite{Lam84}, and in principle protocols could successfully
execute when messages \emph{do not} arrive at certain moments.
Even if we do assume that a fault occurred, aborting may not
be the best choice: sometimes, message retransmission is a better,
more efficient and also more realistic option, as depicted in Figure
\ref{messageflows} $(iv)$.  In this case, a question which arises is
whether to retransmit the original message ($M_{AB}$ for Figure
\ref{messageflows} $(iv)$), or to recompute some parts before
resending the message.  Here, the tradeoff is, as usual, between
efficiency versus security.

Time information can also be included in the contents of $M_{AB}$ and
$M_{BA}$. A typical value to include is a timestamp, to prevent replay
attacks. However, this requires \emph{secure} clock synchronisation of
$A$ and $B$, which is expensive (see Mills~\cite{Mills99} for a
security protocol to achieve this). In fact, this is the reason for
which Bellovin et al.  recommend to switch to nonces in the Kerberos
protocol \cite{BM90}.  Recently, the analysis of security protocols
using timestamps has received considerable attention from the research
community (see Related Work in
Section~\ref{sec:related-work}). Therefore, in this paper we do not
pursue this direction.
\section{A Method for Analysing Security Protocols}
\label{modeling}

We use timed automata~\cite{AD94} to model protocol participants, and
this has several advantages.  Firstly, our method requires the designer
to provide a precise and relatively detailed protocol specification,
which helps to disambiguate the protocol behaviour. Secondly, timing
values like timeouts need to be set at each state, while
retransmissions can be specified as transitions to other
protocol states.  

Once modelled as timed automata, the protocol can be fed to the real
time model checker UPPAAL, which allows the protocol to be simulated
and verified. The simulation provides the designer with a good insight
of the inner workings of protocol, and already at this stage specific
timing values like timeouts can be tuned. Then the designer can
proceed with verification of specific properties. As usual in model
checking, the verification of the protocol with UPPAAL is automatic.

The resulting timed automata model is an informative and precise
description of the security protocol, and thus, it provides a
practical way to strengthen implementations while keeping efficiency
in mind.

As a third and final step we propose to transfer timing information
back to the high level protocol description. This serves to highlight
the role of time in the implementation, but also (as we will
demonstrate in Section \ref{timedchal}), to make timing an integral
aspect of the protocol design.

\subsection{Timed Automata and UPPAAL}

In this paper, the timed automata of Alur and Dill are used for modelling
\cite{AD94}. 
In general, timed automata models have an infinite state space. The
region automaton construction, however, shows that this infinite state
space can be mapped to an automaton with a finite number of
equivalence classes (regions or zones) as
states \cite{AD94}. Finite-state model checking techniques can then be
applied to the reduced, finite region automaton. A number of model
checkers for timed automata is available, for instance,
Kronos \cite{kronos} and UPPAAL \cite{Amn00}.

\begin{figure}[H]
\begin{center}
\epsfig{file=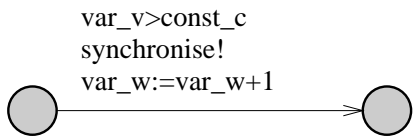, width=6cm}
\end{center}
\caption{Example transition with guard, synchronisation and update}
\label{basic_ta}
\end{figure}

Parallel composition of automata is one of the main sources for
expressiveness. This operation allows to decompose complex behaviour,
thus supporting transparent modelling. When composing automata in
parallel, we need also to provide some form of communication. For the
timed automata we use in this paper, communication comes in form of
hand-shake synchronisation.  Two parallel automata can share a
synchronisation channel, i.e.  both have a transition labelled with a
complementing channel name, e.g. {\it synchronise!}  in the example of
Figure \ref{basic_ta}. These transitions may only be taken together,
at the same moment.  In Figure \ref{basic_ta} we see an example for a
transition, labelled by a guard that has to be true when the
transition is taken, a synchronisation channel, and a variable update.

Data transmission is typically modelled by a synchronisation, where
global variables are updated. These global variables contain the data
that are transmitted.

\begin{figure}[H]
\begin{center}
\epsfig{file=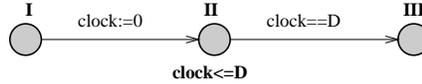, width=6cm}
\end{center}
\caption{Basic timed automaton fragment with a clock and a constant D}
\label{basic_step}
\end{figure}

Timed automata extend ``classical'' automata by the use of real-valued
clock variables.  All clock variables increase with the same
speed. For timed automata we make a difference between a state and a
location: a state is a location where all clocks have a fixed
value. In this sense a location symbolically represents an infinite
set of states, for all the different clock valuations.  In
Figure~\ref{basic_step} an elementary fragment of timed automata is
shown.  When the transition from location {\bf \sf I} to location
{\bf \sf II} takes place, the clock {\it clock} is reset to
0. Location {\bf \sf II} may only be left at time {\bf \sf D}, where
{\bf \sf D} is a constant. The invariant {\it clock$<=$D} at location
{\bf \sf II} enforces that the transition to {\bf \sf III} has to be
taken at time {\bf \sf D}.

Typically, the initial location of an automaton is denoted by a double
circle.  We also make use of committed locations, which have the
property that they have to be left immediately.  In most cases
committed locations are used to model a sequence of actions with
atomic execution.

The properties verified by the model checker of UPPAAL are
reachability properties, like ``there is a state where property {\it
p} holds reachable from the initial state'', or the dual ``in all
reachable states property {\it p} holds''. The latter is falsified if
the model checker finds a state that does not satisfy {\it p}. In this
case a {\it diagnostic trace} from the initial state to the state that
does not satisfy {\it p} is produced by the model checker; it serves as
counterexample.  

We use this mechanism to find attacks. If we can characterize for
example the fact that some secret is not secret any more as a
propositional property, and the model checker finds a state where this
property holds, the diagnostic trace describes a sequence of actions
that leads to this state, which gives precisely the attack.

Note that in this context verification comes very much in the guise of
debugging.  Finding an attack requires an adequate problem model. Not
finding an attack increases the confidence in the modelled protocol,
but does not exclude that attacks could be found in other models for
the same protocol.

\subsection{Overview of the UPPAAL Model}

\begin{figure}
\begin{center}
\setlength{\unitlength}{1973sp}%
\begingroup\makeatletter\ifx\SetFigFont\undefined%
\gdef\SetFigFont#1#2#3#4#5{%
  \reset@font\fontsize{#1}{#2pt}%
  \fontfamily{#3}\fontseries{#4}\fontshape{#5}%
  \selectfont}%
\fi\endgroup%
\begin{picture}(5412,1994)(76,-1283)
\thicklines
{\color[rgb]{0,0,0}\put(2206,269){\oval(210,210)[bl]}
\put(2206,584){\oval(210,210)[tl]}
\put(3496,269){\oval(210,210)[br]}
\put(3496,584){\oval(210,210)[tr]}
\put(2206,164){\line( 1, 0){1290}}
\put(2206,689){\line( 1, 0){1290}}
\put(2101,269){\line( 0, 1){315}}
\put(3601,269){\line( 0, 1){315}}
}%
\thinlines
{\color[rgb]{0,0,0}\put(4081,-331){\oval(210,210)[bl]}
\put(4081,-16){\oval(210,210)[tl]}
\put(5371,-331){\oval(210,210)[br]}
\put(5371,-16){\oval(210,210)[tr]}
\multiput(4081,-436)(234.54545,0.00000){6}{\line( 1, 0){117.273}}
\multiput(4081, 89)(234.54545,0.00000){6}{\line( 1, 0){117.273}}
\multiput(3976,-331)(0.00000,210.00000){2}{\line( 0, 1){105.000}}
\multiput(5476,-331)(0.00000,210.00000){2}{\line( 0, 1){105.000}}
}%
\thicklines
{\color[rgb]{0,0,0}\put(2206,-1156){\oval(210,210)[bl]}
\put(2206,-841){\oval(210,210)[tl]}
\put(3496,-1156){\oval(210,210)[br]}
\put(3496,-841){\oval(210,210)[tr]}
\put(2206,-1261){\line( 1, 0){1290}}
\put(2206,-736){\line( 1, 0){1290}}
\put(2101,-1156){\line( 0, 1){315}}
\put(3601,-1156){\line( 0, 1){315}}
}%
\thinlines
{\color[rgb]{0,0,0}\put(331,-406){\oval(210,210)[bl]}
\put(331,-91){\oval(210,210)[tl]}
\put(1621,-406){\oval(210,210)[br]}
\put(1621,-91){\oval(210,210)[tr]}
\multiput(331,-511)(234.54545,0.00000){6}{\line( 1, 0){117.273}}
\multiput(331, 14)(234.54545,0.00000){6}{\line( 1, 0){117.273}}
\multiput(226,-406)(0.00000,210.00000){2}{\line( 0, 1){105.000}}
\multiput(1726,-406)(0.00000,210.00000){2}{\line( 0, 1){105.000}}
}%
{\color[rgb]{0,0,0}\put(2101,464){\vector( 3, 1){  0}}
\put(2101,464){\vector(-3,-1){1215}}
}%
{\color[rgb]{0,0,0}\put(3601,464){\vector(-3, 1){  0}}
\put(3601,464){\vector( 3,-1){1125}}
}%
{\color[rgb]{0,0,0}\put(4726,-436){\vector(-2,-1){1140}}
}%
{\color[rgb]{0,0,0}\put(826,-511){\vector( 3,-1){1305}}
}%
{\color[rgb]{0,0,0}\put(3076,164){\vector( 0,-1){900}}
}%
\put(2551,389){\makebox(0,0)[lb]{\smash{\SetFigFont{6}{7.2}{\sfdefault}{\mddefault}{\updefault}{\color[rgb]{0,0,0}Intruder}%
}}}
\put(4276,-286){\makebox(0,0)[lb]{\smash{\SetFigFont{6}{7.2}{\sfdefault}{\mddefault}{\updefault}{\color[rgb]{0,0,0}Responder}%
}}}
\put(4276,-886){\makebox(0,0)[lb]{\smash{\SetFigFont{6}{7.2}{\sfdefault}{\mddefault}{\updefault}{\color[rgb]{0,0,0}generate $nb$}%
}}}
\put(4276,-1111){\makebox(0,0)[lb]{\smash{\SetFigFont{6}{7.2}{\sfdefault}{\mddefault}{\updefault}{\color[rgb]{0,0,0}encrypt $nb$,$pk(a)$}%
}}}
\put(2551,-1186){\makebox(0,0)[lb]{\smash{\SetFigFont{6}{7.2}{\sfdefault}{\mddefault}{\updefault}{\color[rgb]{0,0,0}Device}%
}}}
\put(2251,-961){\makebox(0,0)[lb]{\smash{\SetFigFont{6}{7.2}{\sfdefault}{\mddefault}{\updefault}{\color[rgb]{0,0,0}Cryptographic}%
}}}
\put(601,-286){\makebox(0,0)[lb]{\smash{\SetFigFont{6}{7.2}{\sfdefault}{\mddefault}{\updefault}{\color[rgb]{0,0,0}Initiator}%
}}}
\put(4201,389){\makebox(0,0)[lb]{\smash{\SetFigFont{6}{7.2}{\sfdefault}{\mddefault}{\updefault}{\color[rgb]{0,0,0}send/recv}%
}}}
\put(751,389){\makebox(0,0)[lb]{\smash{\SetFigFont{6}{7.2}{\sfdefault}{\mddefault}{\updefault}{\color[rgb]{0,0,0}send/recv}%
}}}
\put(2101,-61){\makebox(0,0)[lb]{\smash{\SetFigFont{6}{7.2}{\sfdefault}{\mddefault}{\updefault}{\color[rgb]{0,0,0}gen nonce,}%
}}}
\put(2101,-286){\makebox(0,0)[lb]{\smash{\SetFigFont{6}{7.2}{\sfdefault}{\mddefault}{\updefault}{\color[rgb]{0,0,0}encrypt,}%
}}}
\put(2101,-511){\makebox(0,0)[lb]{\smash{\SetFigFont{6}{7.2}{\sfdefault}{\mddefault}{\updefault}{\color[rgb]{0,0,0}decrypt}%
}}}
\put( 76,-961){\makebox(0,0)[lb]{\smash{\SetFigFont{6}{7.2}{\sfdefault}{\mddefault}{\updefault}{\color[rgb]{0,0,0}decrypt $\{nb\}_{pk(a)}$}%
}}}
\put( 76,-1186){\makebox(0,0)[lb]{\smash{\SetFigFont{6}{7.2}{\sfdefault}{\mddefault}{\updefault}{\color[rgb]{0,0,0}$\vdots$}%
}}}
\end{picture}
\caption{Structure of our UPPAAL Model}
\label{model}
\end{center}
\end{figure}
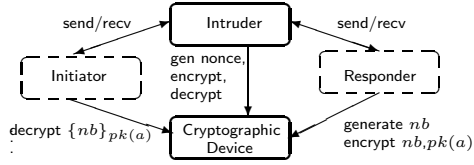

Let us now describe the general form of our model, in some detail. We
model the protocol participants (initiator, responder, etc) and the
intruder as timed automata.  Additionally, we model cryptography as
another automaton, the \emph{cryptographic device}, which acts as an
impartial party that regulates the access to data.  In
Figure~\ref{model}  we illustrate a setting consisting of one
initiator and one responder.
Here, boxes in bold represent our general intruder and the
cryptographic device, while dashed boxes represent the actual
initiator and responder. These participants use the cryptographic
device to perform operations, but communicate through the intruder
(thus the intruder is identified with the network itself, obtaining a
Dolev~Yao like intruder \cite{DY83}).  Our modelling is
\emph{modular}, and allows us to ``plug in'' different participants
(eg., in the analysis of the Yahalom we add a \emph{server}), while
the bold boxes, ie. the intruder and the cryptographic device, are the
core model.

While modelling security protocols as timed automata in UPPAAL,
we will focus on modelling the times required by the principals to encrypt
and decrypt values (and generate nonces), but not on the actual time
that takes the sending (transmission times are assumed to be unknown).
Therefore, for our results to be useful, we assume that computing
times (e.g. cryptographic operations) are not negligible w.r.t.
communication times, and thus choices for timeout values depend both
on communication and computing times.

\subsection{Modelling Cryptography}

\begin{figure}
\begin{center}
\includegraphics{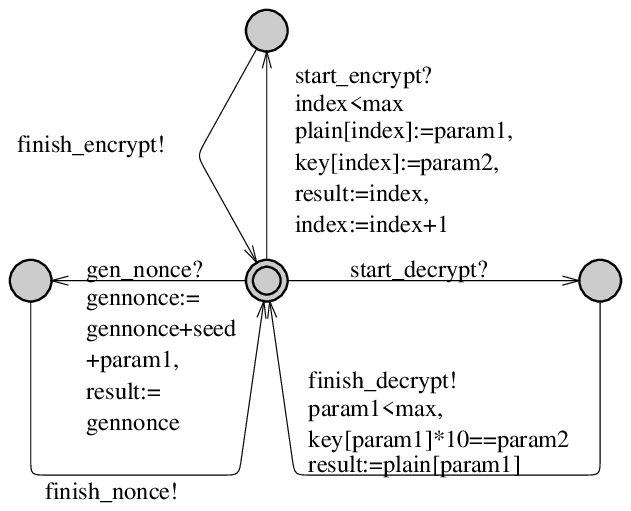}
\caption{Timed automaton for a Cryptographic Device}
\label{figure:device}
\end{center}
\end{figure}

The automaton for a cryptographic device is presented in Figure
\ref{figure:device}. This cryptographic device performs nonce generation and public key cryptography. Later we also use a device for symmetric cryptography, which can be obtained from the one in Figure \ref{figure:device} in a straightforward manner. In fact, our method allows different cryptographic devices to be plugged in as needed (eg. to add hashing).
Basically, the device model is a shared table containing pairs of
plaintexts and keys.  The first service of the cryptographic device is
to provide fresh nonces to the protocol participants (and also the
intruder). The process of nonce generation is started via
synchronization on the $\emph{gen\_nonce}$ channel. To model the new
nonce creation, the local variable \emph{gennonce} is incremented with
the constant \emph{seed} plus the value of {\tt param1} that includes
the ID of the requesting participant (this ensures that initial
generated nonces differ from each other). The number of possible
nonces is bounded by ensuring that \emph{gennonce} is always smaller than a fixed constant
\emph{MAX}. After synchronization, a global {\tt result} variable is
updated with a generated nonce, and the device finishes by
synchronizing on the $\emph{finish\_nonce}$ channel.

Encryption and decryption are modelled by two local arrays to the
cryptographic device, namely {\tt plain} and {\tt key}. When a party
wants to encrypt some value $d$ with key $k$, it synchronises with the
device via the channel $\emph{start\_encrypt}$.  If the device has
still room in its tables, it stores $d$ in the {\tt plain} array and
$k$ in the {\tt key} array. As a result, it sets in the global
variable {\tt result} the \emph{index} in which $d$ and $k$ reside in
the arrays. This index is the ``ciphertext''. Upon decryption, the
ciphertext is provided to the cryptographic device, which then checks
that the provided key is correct: Since we model public-key
encryption, the private key of a public key $k$ is simply modelled as
a function $f$ s.t. $f(k)>MAX$, so that private keys do not clash with
generated nonces and hence are never known by the attacker simply by
guessing.  In this simple case we simply let $f(k)=10k$, which since
only one nonce is needed by the participants in the example protocol of Section~\ref{modeling2},
gives enough room for the attacker to generate nonces.

\paragraph{State constructions}
\begin{figure}
\begin{center}
\includegraphics{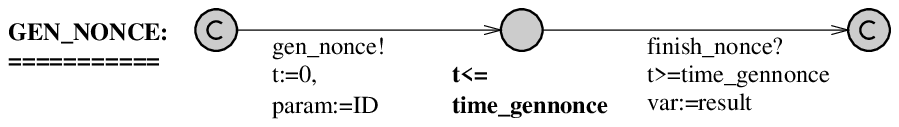}
\includegraphics{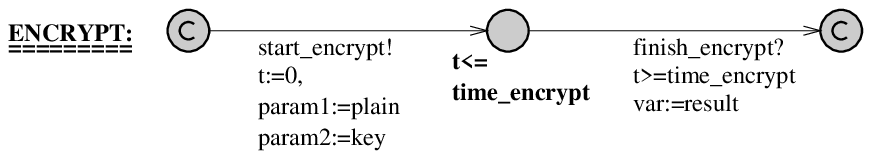}
\includegraphics{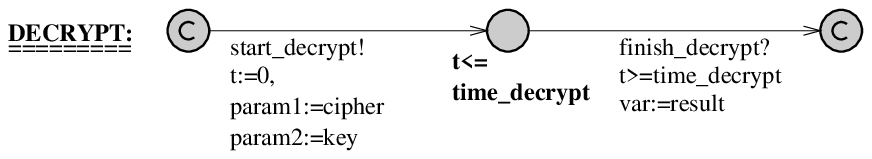}
\caption{State constructions: nonce generation (above), encryption (middle) and decryption (bottom)}
\label{stateconstruction}
\end{center}
\end{figure}

Now that we have the cryptographic device, an honest principal can use
different state constructions to perform cryptographic operations. In
Figure~\ref{stateconstruction} we show the different kinds of state
constructions used in our models, which designers should use as
building blocks for the representation of protocol participants.

In the upper left of Figure~\ref{stateconstruction} we see the building block
for nonce generation. Here, a protocol participant first resets the
clock ${\tt t}$, assigns its identity to variable ${\tt param}$ (used by
cryptographic device to provide different nonces to different
participants) and then fires via the $gen\_nonce$ channel. Then the
participant enters a state in which it waits until the time of nonce
generation happens (${\tt time\_gennonce}$), synchronises via the
$finish\_nonce$ channel and obtains the return value via variable ${\tt
  result}$. Encryption and decryption are analogous, and only differ
in that they use two parameters ${\tt param1}$ and ${\tt param2}$ (for
plaintext and key in the former, and ciphertext and key in the
latter).

\subsection{Modelling the Adversary}
\label{sec:modelling-adversary}
\begin{figure*}
\begin{center}
  \includegraphics{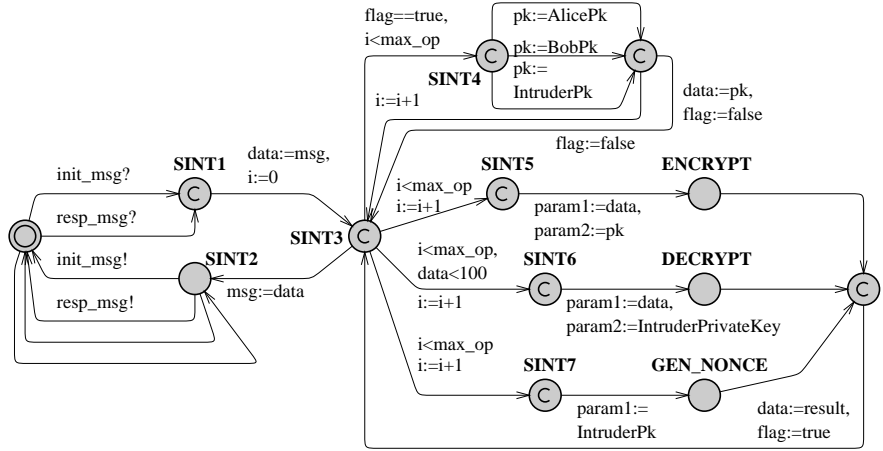}
\caption{Schema for the timed automaton for the Intruder}
\label{figure:intruder}
\end{center}
\end{figure*}

The intruder, presented in Figure \ref{figure:intruder}, works
basically as a Dolev~Yao intruder \cite{DY83}. The intruder models the
network itself, by acting as an intermediary of communication between
the initiator and responder. This is modelled by letting the intruder
synchronise on both channels $init\_msg$ and $resp\_msg$.  Upon
synchronising by receiving a message, the intruder moves to state
(SINT1), where it saves the message ${\tt msg}$ in its local variable
${\tt data}$ and resets an index variable $i$ which bounds the total
number of actions allowed to do before continuing execution. Then, the intruder
moves to state (SINT3), where it makes a nondeterministic choice for an
action. More precisely, it can decide to:\\
-- Choose an identity in its local variable ${\tt pk}$ (State SINT4)\\
-- Encrypt a value (State SINT5)\\
-- Decrypt a value (State SINT6)\\
-- Generate a nonce (State SINT7)\\
-- Save  variable {\tt data} as message {\tt msg}.\\
The intruder can then continue to perform these actions, choose to send a
message or simply block a message and continue the execution.
Moreover, the intruder can also delay arbitrarily a message, by
waiting in state (SINT2).

Note that the intruder is independent of the actual protocol under
study, and hence it is generic to analyze protocols using public key
encryption (although this intruder is not able to concatenate
messages; we extend it in the next section).

\section{Analysing Protocols}
\label{sec:analysing-protocols}
We first consider a simple protocol to illustrate our technique.
Later, we move on to analyse the more complex Yahalom protocol.

\subsection{An Example Protocol}
\label{modeling2}
In this section we study and model in UPPAAL a simplified version of
the Needham Schroeder protocol, thoroughly studied in the literature
(see eg. \cite{Lowe95}).
Differently from the Needham Schroeder protocol whose goal is to
achieve mutual authentication, our simpler protocol aims at
authenticating the initiator $A$ to a responder $B$ only (we do not
lose generality here, this is just a simplification to improve
presentation).  The protocol is as follows:
\begin{eqnarray*}
1. \ A \rightarrow B &:& A\\
2. \ B \rightarrow A &:& \{N_B\}_{K_A}\\
3. \ A \rightarrow B &:& \{N_B\}_{K_B}
\end{eqnarray*}

In the first message, the initiator $A$ sends a message containing its
identity to the responder $B$. When $B$ receives this message, it
generates a nonce $N_B$, encrypts it with the public key $K_A$ of $A$
and sends it back to $A$. Upon receipt, $A$ decrypts this message with
her private key, obtains the nonce $N_B$, reencrypts it with the
public key $K_B$ of $B$ and sends it back to $B$.


We can now move on to describe the actual initiator, responder and
intruder. Both the initiator and responder have local constants {\tt
  time\_out}, which represent their timeout values. Also, the
initiator, responder and intruder have local constants {\tt
  time\_gennonce}, {\tt time\_encrypt} and {\tt time\_decrypt} that
represent the time required to generate a nonce, encrypt a value or
decrypt a value, respectively for \emph{each} principal. 

\begin{figure*}
\begin{center}
  \includegraphics{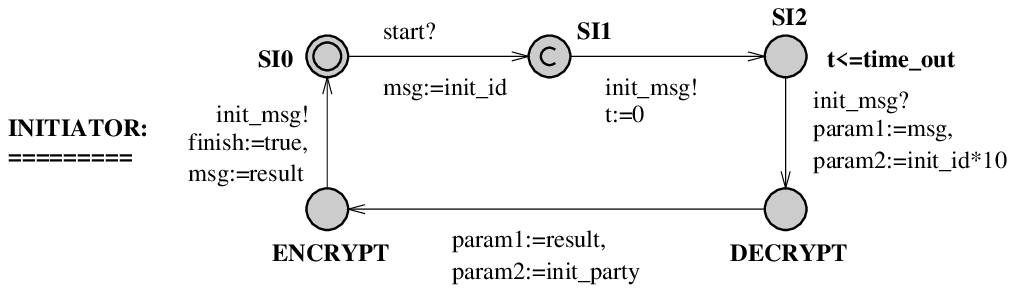}
\includegraphics{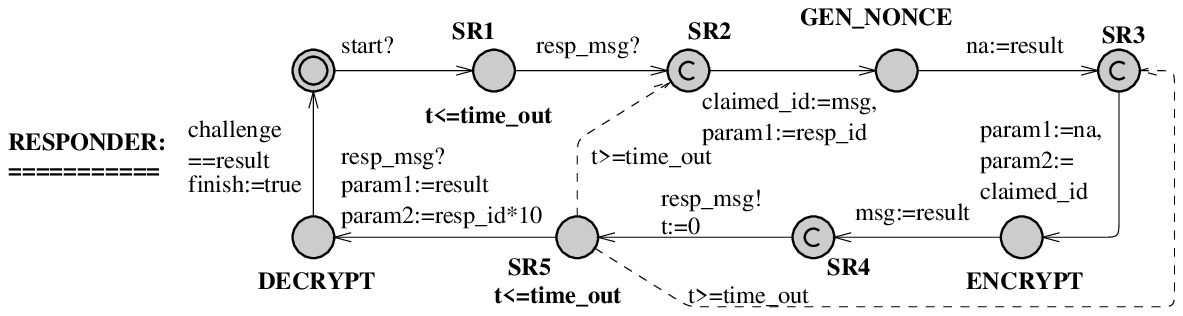}
\caption{Schemas of timed automata for the Initiator (top) and the Responder (bottom)}
\label{figure:parties}
\end{center}
\end{figure*}

The automata for the initiator and responder of our simple protocol
presented above are given in Figure \ref{figure:parties} (the dashed
transitions of the responder correspond to retransmissions, discussed
in Section \ref{retrans}).  The initiator $A$ starts her execution
when activated via channel $start$ (State SI0).  The actual identity of
 the initiator role is set via the global variable {\tt
  init\_id} (this and other role variables are chosen by the {\tt
  Init} automaton, described below). The initiator saves {\tt
  init\_id} as the first message (see protocol message 1).  Then, the
initiator starts her protocol execution, by firing via the channel
$init\_msg$.
After this, the initiator starts a clock $t$ and waits for a response,
or until $t$ reaches {\tt time\_out} (State SI2). If the timeout
occurs, the protocol is aborted (a retransmission at this point would
be equivalent to restart the protocol). If a response is received
before the timeout via the $init\_msg$ channel, $A$ tries to decrypt the
received message {\tt msg}.  This takes time {\tt time\_decrypt} for
the initiator.  After the decryption, the initiator reencrypts the
obtained nonce (stored in {\tt result}) and finally sends
the last message via the $init\_msg$ channel, setting to {\tt
true} its local boolean variable {\tt finish}.

The responder automaton $B$ works similarly to the initiator. After
receiving the start signal, $B$ waits for the message containing the
claimed identity of $A$ (State SR1).  When received, $B$ saves the
first message in the local variable {\tt claimed\_id}. After this, $B$
generates a nonce by contacting the cryptographic device. When ready
(State SR3), $B$ encrypts the nonce with the value received in Message
1 (we identify identities with public keys). After finishing the
encryption (State SR4), the message is sent and $B$ starts to wait for
a response (State SR5). If an answer comes before the timeout, $B$
decrypts the message and checks that the challenge is indeed the one
$B$ sent.  If so, the local boolean variable {\tt finish} is set to
{\tt true}.



\subsubsection{Verification}
\begin{figure}
\begin{center}
\includegraphics{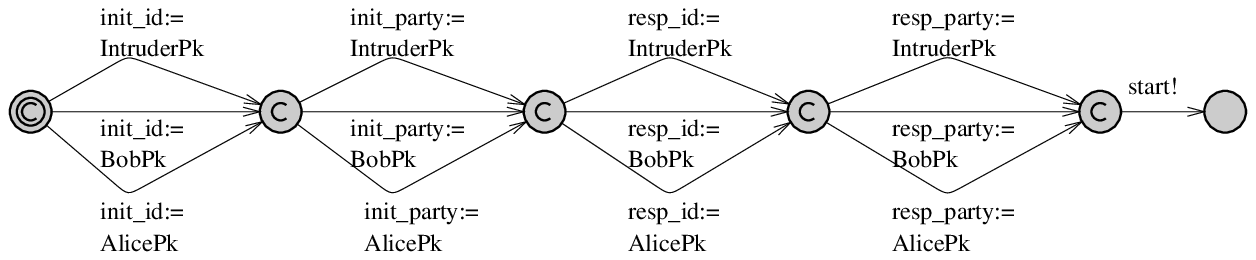}
\caption{Timed automaton for the Init automaton}
\label{figure:init}
\end{center}
\end{figure}

We wish to verify that our simple protocol indeed accomplishes
authentication of $A$ to $B$. To this end, we will model check one
session of the protocol containing one initiator, one responder and
one intruder. We use a special $Init$ automaton that instantiates the
initiator and responder with identities (like $A$, $B$ and $I$), and
then starts the execution run by broadcasting via the $start$ channel.
 The init
automaton is given in Figure \ref{figure:init}. 

The property we check, $AUT$, is shown in Table 1.
\begin{table*}
\begin{center}
\begin{tabular}{|lll|}
\hline
$AUT$&$=$&${\tt E<> Responder.finish \ and }$\\
&&${\tt \ Responder.claimed\_id!=resp\_party}$\\
$AUT_y$&$=$&${\tt E<>\ Initiator.finish\ and\ }$\\
&&${\tt Initiator.ticks<(Responder.time\_encrypt}$\\
&&${\tt +Responder.time\_gennonce+Server.time\_encrypt*2}$\\
&&${\tt +Server.time\_decrypt)-1}$\\
\hline
\end{tabular}
\end{center}
\caption{UPPAAL properties}
\label{auttable}
\end{table*}
$AUT$ states that if we reach a state in which the responder has
finished executing but the claimed id (corresponding to the first
message of protocol) does \emph{not} coincide with the actual identity
of the initiator,  the protocol is flawed. Indeed, a state in
which the initiator can ``lie'' and still force the responder to
finish means that authentication is violated. 
This is one of the possible forms of authentication failure. It is
  outside the scope of this paper to illustrate different
  authentication flaws (see Lowe \cite{Lowe97} and Cremers et
  al.~\cite{cmv03} for more on authentication notions).

If we use a long timeout for $B$, i.e.
\begin{small}
\[
B.{\tt time\_out}\geq
Intruder.{\tt time\_decrypt} + Intruder.{\tt time\_encrypt} + 
A.{\tt time\_encrypt} + A.{\tt time\_decrypt}
\]
\end{small}
here UPPAAL finds a
man-in-the-middle attack, presented on the left hand side of Table
\ref{table}.
\begin{table*}
\begin{center}
\begin{tabular}{|lrl|rll|}
\hline
$\alpha.1$\  &$ A \rightarrow I$& :$ A$&$1. \ I(B) \rightarrow B$ &:&$ B$\\
$\beta.1$\  &$ I(A) \rightarrow B$& :$ A$&$2. \ B \rightarrow I(B) $&:& $\{N_B\}_{K_B}$\\
$\beta.2$\  &$ B \rightarrow I(A)$& : $\{N_B\}_{K_A}$&$3. \ I(B) \rightarrow B$&:& $\{N_B\}_{K_B}$\\
$\alpha.2$\  &$ I \rightarrow A$&: $\{N_B\}_{K_A}$&&&\\
$\alpha.3$\  &$ A \rightarrow I$&: $\{N_B\}_{K_I}$&&&\\
$\beta.3$\  &$ I(A) \rightarrow B$& : $\{N_B\}_{K_B}$&&&\\
\hline
\end{tabular}
\end{center}
\caption{A man-in-the-middle attack (left) and a replay attack (right)}
\label{table}
\end{table*}
This attack is similar to Lowe's attack \cite{Lowe95}, in which an
attacker fools $B$ into thinking he is communicating with $A$, while
in reality $A$ only talks to $I$. Of course, we could patch the
protocol as Lowe did. But, in the context of time, it is interesting
to model-check the protocol with a \emph{tighter} timeout, ie.
$B.{\tt time\_out}<Intruder.{\tt time\_decrypt}+Intruder.{\tt
time\_encrypt}+A.{\tt time\_encrypt}+A.{\tt time\_decrypt}$. When this
constraint is verified, the man-in-the-middle attack vanishes. Of
course, we cannot pretend that $B$ knows the intruder's times of
encryption and decryption. 
Nevertheless, $B$ can set $B.{\tt time\_out}= A.{\tt
  time\_encrypt}+A.{\tt time\_decrypt}$, leaving \emph{no} space for
any interruption.

A second attack which is independent of timeouts (even if we set
$B.{\tt time\_out}=0$!) was also found by UPPAAL; this time,
the vulnerability is much simpler. We report it on the right hand side
of Table \ref{table}.  This attack corresponds to a ``reflection''
replay attack~\cite{Syv94}. This attack occurs when the intruder simply
replies $B$'s message. The attacker fools $B$ into thinking its
communicating with himself, while it is not true in reality.
Interestingly, suppose we change message 3 of the protocol to $3'.\
A\rightarrow B: \{N_B+1\}_{K_B}$. Now, the above replay attack is
prevented, since message 2 is not valid as message 3 anymore.
Of course, a patch \`a la Lowe
for \emph{both} also prevents both problems:
\begin{eqnarray*}
1. \ A \rightarrow B &:& A\\
2. \ B \rightarrow A &:& \{B,N_B\}_{K_A}\\
3. \ A \rightarrow B &:& \{N_B\}_{K_B}
\end{eqnarray*}

Having find confirmation that our framework is capable of
finding \emph{untimed} attacks (and thus confirming known attacks), we
proceed to provide a good baseline to study extended security protocols
with timing issues, like timeouts and retransmissions.

\subsubsection{Retransmissions}
\label{retrans}
Consider again the automaton for the responder, given in Figure
\ref{figure:parties}. In state SR4, the responder sends the
challenge $\{N_B\}_{K_A}$, and waits for a response in state SR5.
If the response does not arrive before the timeout, the responder
simply aborts. Now we consider possible retransmissions that allow the
protocol to recover and continue its execution.  With timed automata,
retransmissions are easy to model by adding transition arrows from
state SR5 to previous states of the automaton (the dashed lines in
Figure \ref{figure:parties}); These transitions are guarded, allowing
to perform the action only when the timeout is reached (ie., ${\tt
  t>=time\_out}$).  
A further refinement not explored here would be to add counters so
that the number of retransmissions can be limited before aborting.

We consider two potential target states for the timeout of the
Responder in SR5, namely states SR3 and SR2. Choosing the former
corresponds to retransmitting the exact same message that was sent
before, $\{N_B\}_{K_A}$. On the other hand, linking the retransmission
arrow to SR2 corresponds to recomputing the whole message,
by creating a new nonce $N'_B$ and sending $\{N'_B\}_{K_A}$.

We implemented both strategies in our UPPAAL model. As can be
expected, retransmitting the exact message \emph{once} has the effect
of \emph{duplicating} the timeout for $B$, and thus the
man-in-the-middle attack becomes possible even for tight timeout
values. On the other hand, recomputing the whole message preserves the
security of the protocol, at a higher computational cost.  This
evidences that indeed these design decisions are important for both
security and efficiency, and a careful analysis can help to choose the
best timeouts and retransmissions for a practical implementation.

\subsection{A Real Protocol}
\label{sec:real-protocol}
Having illustrated our approach with a simple example we now study a
more realistic protocol, the Yahalom protocol \cite{CJ96}. This
protocol aims at authentication of $A$ and $B$ as well as key
distribution of $A$ and $B$ using a shared server $S$ with whom both
$A$ and $B$ share secret keys $K_{AS}$ and $K_{BS}$.

Our choice
is based on the fact that Yahalom is a complex and strong protocol,
with no known attacks over it (However, a modification proposed by
Abadi et al.~\cite{BAN90} has a known type-flaw attack). Our aim is to
study the protocol in more detail (and thus closer to an
implementation) with timing information.  The protocol is as follows:
\begin{eqnarray*}
1. \ A \rightarrow B &:& A,N_A\\
2. \ B \rightarrow S &:& B,\{N_B,A,N_A\}_{K_{BS}}\\
3. \ S \rightarrow A &:& \{B,K_{AB},N_A,N_B\}_{K_{AS}},\{A,K_{AB},N_B\}_{K_{BS}}\\
4. \ A \rightarrow B &:&\{A,K_{AB},N_B\}_{K_{BS}},\{N_B\}_{K_{AB}}
\end{eqnarray*}

Here we use symmetric encryption, and key $K_{XY}$ is shared between
$X$ and $Y$.

\begin{figure}
\begin{center}
\includegraphics{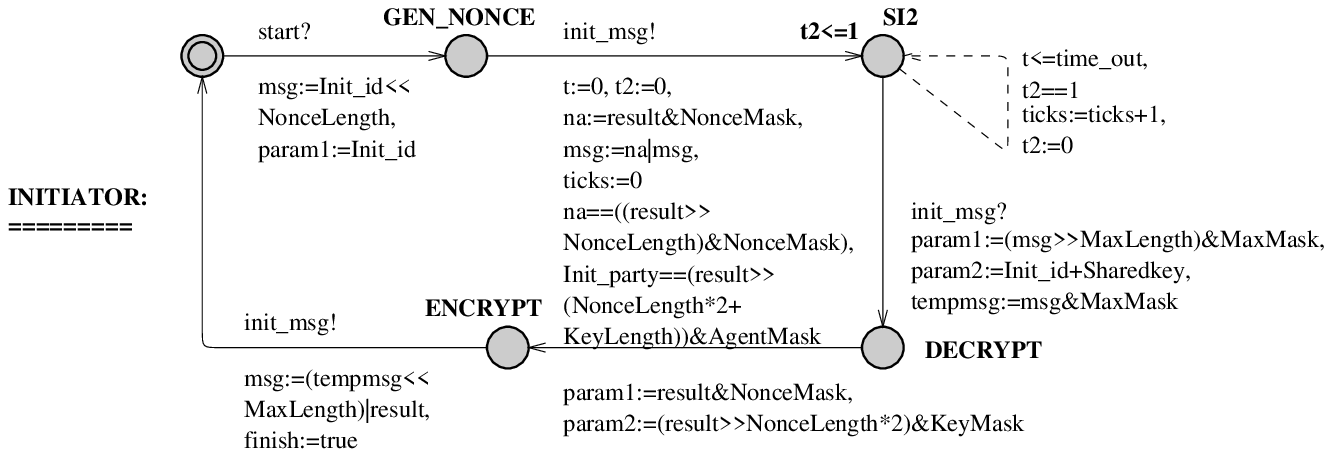}
\includegraphics{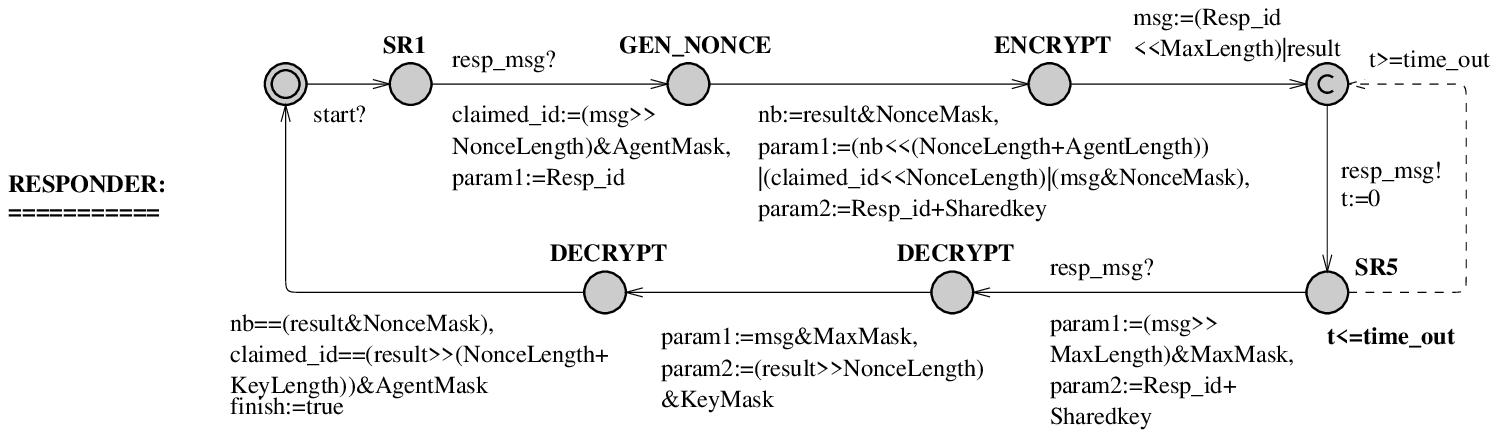}
\caption{Timed automata schemas for the Yahalom Initiator (top) and responder (bottom)}
\label{inityahalom}
\end{center}
\end{figure}

\begin{figure}
\begin{center}
\includegraphics{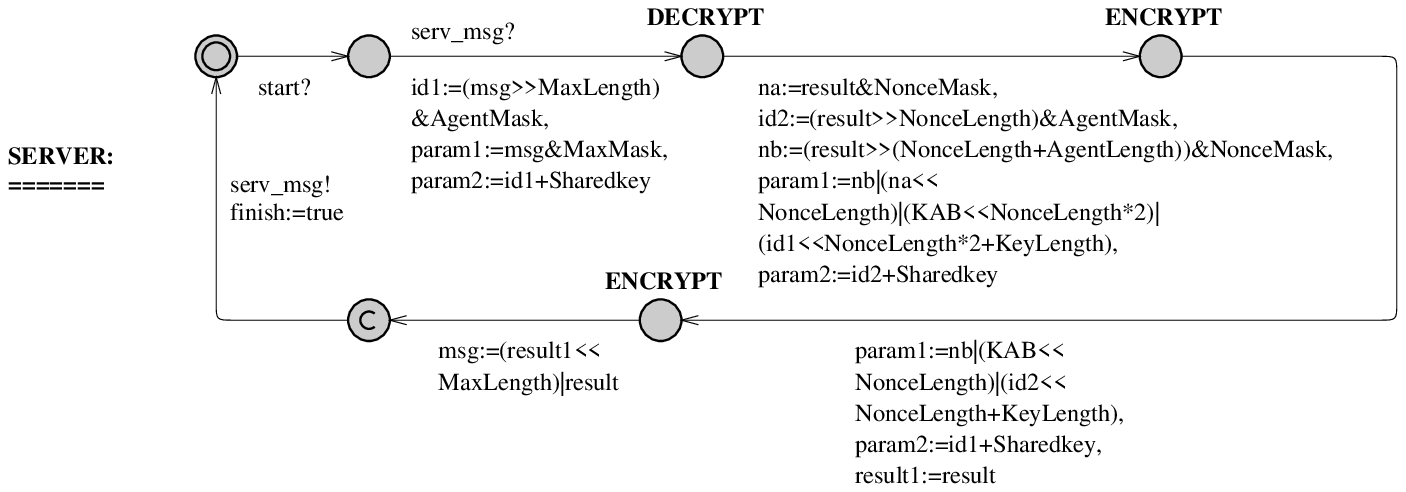}
\includegraphics{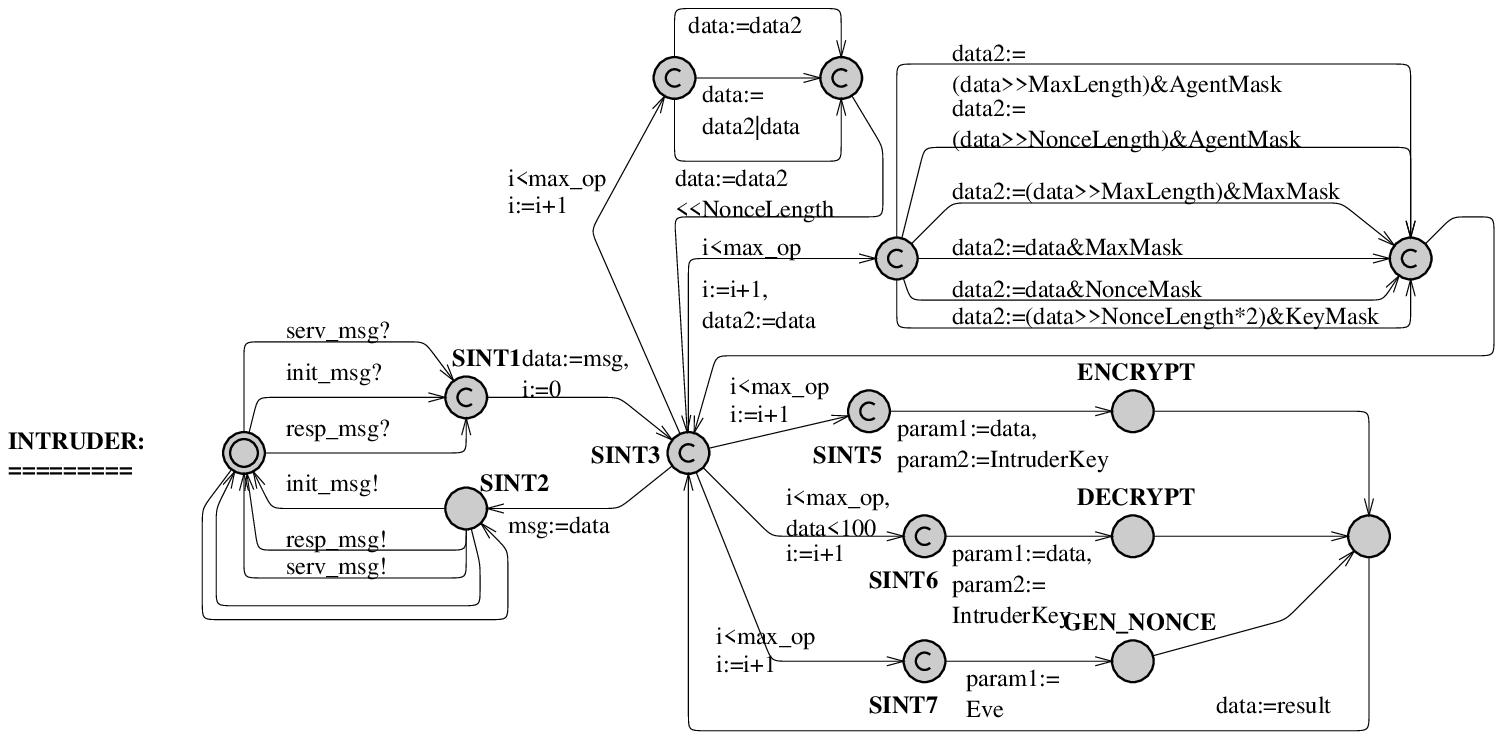}
\caption{Timed automata schemas for the Server (top) and Intruder (bottom)}
\label{inityahalom2}
\end{center}
\end{figure}

To model concatenation in an efficient way, we gathered several
message components into a 16 bit field, thus keeping the state space
as small as possible. In our case, we assume that nonces have $4$
bits, principal id's $2$ bits and keys $4$ bits. To access these values,
we use bit-wise {\tt and} with appropriate masks, and (left,right) bit
shifts.  Our intruder has also the capability to do the shifts and
mask, and we also removed the ``public key'' choice from the intruder
of Figure \ref{figure:intruder}.
We have modelled the protocol in UPPAAL (the initiator, responder,
server and intruder are shown in Figures
\ref{inityahalom} and \ref{inityahalom2}).


As we did with the previous protocol, first we check whether
authentication of $A$ to $B$ could be falsified, using property $AUT$
from Table \ref{auttable}.  This property is not satisfied, confirming
that Yahalom is secure.  Now we move to study time sensitive issues.

There are two places in which timeouts and retransmissions can occur
in this protocol.
The first one is in Message 1: After $A$ sends her message, she
starts a timer waiting for message 3. Now, suppose that a timeout
occurs, and $A$ wants to retransmit her message. We can be confident
that resending the same nonce $N_A$ will \emph{not} affect security,
since in any case it was already sent in the clear in the first time.
However, an interesting timing issue arises here. An answer that is
received \emph{too early} by $A$ could be suspicious, because some
time must pass while $B$ talks to $S$. If $A$ knows $B$'s and $S$'s
encryption and decryption times, $A$ could even deliberately
``hibernate'' (eg.  to save energy) until the response is likely to
arrive (this models a windowed timeout, see Figure~\ref{messageflows}
$(ii)$).  We model checked this property by measuring the time after
$A$ sends her message, and a response arrive (we count {\tt ticks},
the dashed loop transition of the initiator in Figure
\ref{inityahalom}).  The specified property is $AUT_y$, shown in Table
\ref{auttable}.  This property is not satisfied, confirming that there
is no way that the initiator can receive a valid answer \emph{before}
the time required by the responder and server to process $A$'s
request.  In an implementation, it is reasonable for $A$ to set a
timeout like above, since it is realistic to assume that $A$ can know
the responder and server's times of encryption and decryption.

The second timeout is set by $B$ after sending his message at step 2.
If a timeout occurs, the retransmission decision is more delicate: It
is not clear whether $B$ should resend the original message, should
recompute $N_B$ or whether $B$ should abort, since clearly $N_A$ cannot be
recomputed.  Intuitively, $N_B$ could be reused.
We modelled in UPPAAL the retransmission of the exact message (as the
dashed
transition of the responder in Figure \ref{inityahalom}). 
When we model check again property $AUT_y$, we obtain that it is still
unreachable, confirming that in that case an efficient retransmission
of the same message 2 by $B$ is secure.

\begin{figure}
\begin{center}
\input{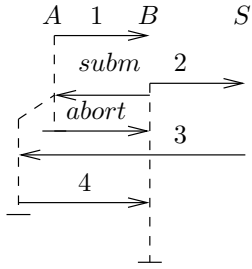}
\caption{An more detailed implementation of Yahalom}
\label{yahalom}
 \end{center}
\end{figure}

However, by observing the messages flow, we know that if $B$ timeouts
then it is very likely that $A$ has also reached its timeout and
aborted (see Figure \ref{messageflows} $(iv)$). This mainly happens
because since $A$ is unsure whether $B$ is alive or not, and thus
$A$'s timeout needs to be tight. If $A$ knew that $B$ is alive and
waiting for an answer from $S$, then $A$ could \emph{extend} its
timeout. We then sketch a more efficient implementation in
Figure~\ref{yahalom}, in which at Message 2 $B$ also sends a special
$subm$ message notifying $A$ of the submission to $S$. Then $A$ can
extend its timeout with more confidence (the second dashed line in
Figure \ref{yahalom}). In the case $subm$ is never received by $A$,
she can send an \emph{abort} message to $B$. Of course, in this simple
model the attacker can also send this messages, thus performing denial
of service attacks; in any case, our attacker is powerful enough to
stop communication altogether.

In summary, for the Yahalom protocol we obtain that retransmitting for the
responder is secure, and also that the initiator can be implemented to
efficiently ``hibernate'' a safe amount of time before receiving a
response. 
  
%



\section{Taxonomy of Message Flows in Security Protocols}
\label{sec:taxon-mess-flows}

\begin{figure}[!h]
\begin{center}
\input{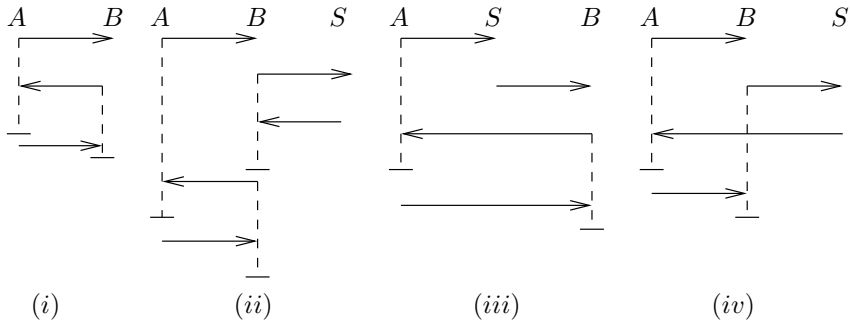}
\caption{Typical message flows for authentication protocols}
\label{messageflows2}
\end{center}
\end{figure}

The flow of messages of many protocols follow a small set of specific
patterns.  By exploring the well known Clark Jacob library \cite{CJ96}
of authentication protocols and the Security Protocols Open Repository
(SPORE) \cite{spore}, we were able to categorize the protocols in four
categories, as shown in Figure \ref{messageflows2}. To each pattern,
we add the corresponding timeouts, and analyze their impacts on
security and efficiency. For the original references of the protocols,
the interested reader may consult the Clark Jacob library \cite{CJ96}
and the SPORE library \cite{spore}.

Not shown in this categorization are non-interactive protocols which
do not wait for messages and thus do not require timeouts. In this
category fall the Wide Mouthed Frog protocol, the CCITT X.509
simple pass protocol and the CAM protocol for mobile IP.

First, we discuss the simplest pattern in Figure \ref{messageflows2}
$(i)$. This is a three-message exchange with two participants.  This
pattern is the simplest and also the most secure one from timing point
of view, since timeouts can be set tight, due to the ping-pong nature
of the exchanges. To this pattern correspond both the example protocol
of Section \ref{modeling2} and the one in Section \ref{timedchal}, and
also the protocols CCITT.509 three pass, the Shamir Rivest and Adleman
Three-Pass protocol, the ISO XXX Key Three-Pass (and their repeated
protocols), the SmartRight view-only protocol (from SPORE) and the
Diffie-Hellman key exchange protocol. With a fourth message from $B$
to $A$ in the same fashion we find the Andrew Secure RPC
protocol. Adding a third participant $S$, but still doing ping-pong
exchanges, we can add the Needham Schroeder symmetric key protocol and
the Denning Sacco protocol.

Secondly, we identify three-party protocols, in which a server $S$
also takes part in the communication (Figure \ref{messageflows2}
$(ii)$) but ping-pong exchanges are not anymore used.  This pattern is
potentially unsafe and inefficient for $A$, since she has to wait
until a long timeout as elapsed after the first message before
receiving an answer from $S$. This is due to the fact that three
messages have to be exchanged after $A$'s initial message. By
consulting again the Clark and Jacob library and the SPORE repository,
we see that the Otway Rees protocol, the Gong mutual authentication
protocol, the Woo-Lam mutual authentication protocol, the Carlsen
protocol, and finally the Kehne Schoenwalder Langendorfer (KSL)
protocol all fall in this category.  Adding ping-pong exchanges before
and after the exchanges of Figure \ref{messageflows2} $(ii)$ we find the
Needham Schroeder Public key protocol. Adding a ping-pong exchange
before Figure \ref{messageflows2} $(ii)$, and removing the last exchange
gives us the SPLICE/AS protocol.

Thirdly, we see a pattern to which only the Kao Chow protocol belongs
in Figure \ref{messageflows2} $(iii)$. This pattern is however better
than $(ii)$, since shorter and fewer timeouts are used: $A$ needs to
wait for the timeout corresponding to only two messages (instead of
three as in $(ii)$), and $B$ has to wait for only one timeout
(comparing to two timeouts in $(ii)$).

Finally, in Figure \ref{messageflows2} $(iv)$ we see the last pattern.
This pattern is worse than $(iii)$ since $B$ needs to wait longer (for
two messages instead of one in $(iii)$). However, it is unclear
whether it is better than $(ii)$, which uses two timeouts of one
message each: the actual efficiency and security depends on the actual
timeout values used in each case.  This category is inhabited by the
Yahalom and Neuman Stubblebine protocols.

This taxonomy shows how authentication protocols can be categorized
into a handful of patterns. The efficiency and security that an
implementation of a protocol will have depends on which pattern the
protocol follows, and thus it is useful to consider the patterns in
isolation from the actual protocols. In this paper we do not pursue
this further, although as future work it would be interesting to
further analyze these abstract timing patterns induced from security
protocols.

\section{Beyond Model Checking: Novel Issues Considering Time}
\label{issues}
So far our method has been used for analysis purposes, ie. to
model, classify and debug security protocols as a source of hints for the
improvement of the protocol implementations. We now explore some ideas
to improve the protocols themselves, and also present the threat of a
more subtle attack, based only on timing.


\subsection{Using Time as Information: Timed Challenges}
\label{timedchal}

Sometimes it can be useful to include other timed information than
timestamps, \emph{even} if the clocks are not synchronised.  Consider
the following protocol, obtained by omitting the encryption of the
last message of the (patched) protocol of Section \ref{modeling2}:
\begin{eqnarray*}
1. \ A \rightarrow B &:& A\\
2. \ B \rightarrow A &:& \{B,N_B\}_{K_A}\\
3. \ A \rightarrow B &:& N_B
\end{eqnarray*}

Even though $N_B$ is now sent in the clear, this protocol still
achieves authentication of $A$ to $B$, although now the nonce
obviously cannot be regarded as a shared secret. Still, the intruder
can prevent a successful run of the protocol (eg. by intercepting
message 3), hence the protocol is as strong as it was before in this
respect.

Imagine now a situation in which there is a link from $A$ to $B$ in
which data can be sent very fast, but at a high cost per bit sent.
For example, think that the high cost of sending information comes
from the fact that we have devices with a very limited amount of
energy, like wireless sensor networks for instance.  Alternatively, in
some networks, operators charge according to quality of service, and
many networks have asymmetric links (eg. Cable modem and ADSL).

Assume therefore that, sending $N_B$ in message 3 is expensive and not
desirable.  We propose a solution based exclusively on
using \emph{time} as information.  Let $\delta_{AB}$ be the average
time it takes for a message to be sent from $A$ to $B$, and
analogously $\delta_{BA}$. Then consider the ``timed'' variant of the
above protocol, demonstrating how timing information is brought back
to the (abstract) protocol level (ie. Step 3 of
Section \ref{modeling}):
\begin{eqnarray*}
1. \ A \rightarrow B &:& A\\
2. \ B \rightarrow A &:& \{B,t_B\}_{K_A}\\
3. \ A \rightarrow B &:& ``ack\text{''} \text{ at time }t_B-\delta_{AB}-\delta_{BA}
\end{eqnarray*}

In Message 2, instead of a nonce, $B$ generates some random time value
$t_B>\delta_{AB}+\delta_{BA}$, concatenates it with $B$'s identity and
encrypts the message with $A$'s public key. Then, $B$ starts a timer
$t$ and sends the message.  Upon reception, $A$ extracts $t_B$,
\emph{waits} time $t_B-\delta_{AB}-\delta_{BA}$, and replies the
single bit message ``$ack$''. When $B$ receives this message, the
timer is stopped and $B$ checks that $t$ is \emph{sufficiently} close
to $t_B$; if so, $A$ is authenticated to $B$.  Of course, the amount
of noise in the time measurements influences what we mean by
``\emph{sufficiently} close'' above.  Also, to be realistic, the
length in bits of $t_B$ should be small enough, otherwise $B$ would be
waiting too long (on average); this would give an intruder the chance
to guess $t_B$, and answer the ``$ack$'' at the appropriate
time. Moreover, if encryption $\{\cdot\}_\cdot$ is deterministic, an
attacker can record $\{B,t_B\}_{K_A}$ and the answers $t_B$ in a
table. As soon as $A$ chooses $t_B$ again (and this is likely since
$t_B$ is small) and sends out $\{B,t_B\}_{K_A}$, the attacker can
notice the repeated message in its table since encryption is
deterministic, and hence the attacker can intercept the message (so it
never arrives to $B$) and still authenticate to $A$, since the
attacker knows precisely when to send $t_B$. Probabilistic encryption
solves this issue, although even then the attacker can simply guess
$t_B$ and violate authentication with non-negligible
probability. However, we can strengthen the protocol as follows:
\begin{eqnarray*}
1. \ A \rightarrow B &:& A\\
2. \ B \rightarrow A &:& \{B,t_{B_1},\hdots,t_{B_n}\}_{K_A}\\
3. \ A \rightarrow B &:& ``ack\text{''} \text{ at time }t_{B_1}-\delta_{AB}-\delta_{BA}\\
\ \  \ \ \ \ \ \vdots \ \ \ \ & &\\
n+3. \ A \rightarrow B &:& ``ack\text{''} \text{ at time }t_{B_n}-\delta_{AB}-\delta_{BA}
\end{eqnarray*}

For example, if $t_{B_i}$ is of length $4$ bits, for $i\in[1..n]$,
then the total answer is $n$ bits, in comparison with an answer of
$4n$ bits required in the nonce protocol. 

Of course, sending several short messages can be worse than sending
one long message, in which case our protocol would not be so useful.
In general, the value of $n$ must be chosen as small as possible,
depending on the desired security and network latency. A fast network
allows us to reduce $n$ and at the same time increment the length of
$t_{B_i}$, for $i\in[1..n]$.

Intuitively, the sent times of the ``$ack$'''s represent information,
and the above protocols exploit that. To the best of our knowledge,
this is a novel usage of time in security protocols.

\paragraph{Application} 
This protocol can be used to authenticate a whole chain of network
packets, as follows.  Suppose $A$ has a large sequence of $n$ packets
which must be streamed to $B$ over a network. For instance, these
packets can represent an audio stream in the Internet.  We want to
authenticate the audio stream, but we do not wish to spend lots of
resources on doing this. Let
$t_{B_i}\in \{\delta_{AB}+\delta_{BA},\delta_{AB}+\delta_{BA}+C\}$ for
some constant $C$ and $p_i$ denote packet $i$, for $1\leq i \leq
n$. Then the protocol becomes:
\begin{eqnarray*}
1. \ A \rightarrow B &:& A, n\\
2. \ B \rightarrow A &:& \{B,t_{B_1},\hdots,t_{B_n}\}_{K_A}\\
3. \ A \rightarrow B &:& p_1 \text{ at time }t_{B_1}-\delta_{AB}-\delta_{BA}\\
\ \  \ \ \ \ \ \vdots \ \ \ \ & &\\
n+3. \ A \rightarrow B &:& p_n \text{ at time }t_{B_n}-\delta_{AB}-\delta_{BA}
\end{eqnarray*}

When $t_{B_i}$ is $\delta_{AB}+\delta_{BA}$, it is the delay
introduced by $A$ is zero, ie. $p_i$ is sent right away. However, when
$t_{B_i}$ is $\delta_{AB}+\delta_{BA}+C$, the delay is $C$. To be as
efficient as possible, $C$ should be chosen to be the minimum amount
of time that allows $B$ to distinguish the delay modulated by $A$.

 In this protocol, only one bit is authenticated per packet. However,
the larger the $n$ is, the more confidence we can obtain of $A$'s
authentication.

\paragraph{Discussion} In this protocol, we are in reality exploiting a well-known feature of channel coding: a so-called \emph{timing covert channel}. In such a channel, the transmitting party modulates packets so that information can be passed even if its not allowed by the environment. Our usage differs in three ways:
\begin{itemize}
\item Firstly, we use a mixed approach, in which some information is sent in the standard channel, and other is sent in the timing channel.
\item A second difference is more fundamental than the previous one. Our usage of the timing channel is purposely public, and there is no environment trying to stop the unauthorized information flow. Timing is used only because of its practical advantages, namely low-bandwidth.
\item Finally, in our protocol both communicating parties do not
  initially trust on the other's identity, in principle:
  Indeed, ours is an \emph{authentication} protocol.
\end{itemize} 

\subsection[Timing Leaks]{Timing Leaks in an Implementation of the Private Authentication Protocol}
\label{section:private}
We now present a threat against an implementation of security
protocols with branching: the so called timing attack. We illustrate
this by showing an attack over a careless implementation of Abadi's
Private Authentication (PAP) protocol~\cite{Aba00} (The \emph{second}
protocol). It is worthwhile to mention that the protocol has been
proved correct by Abadi and Fournet~\cite{FA02}, in a setting without
time.
We assume that each principal $X$ has a set of communication parties
$S_X$, listing the principals with whom $X$ can communicate.  The aim of
the protocol is to allow an principal $A$ to communicate \emph{privately}
with another principal $B$. Here ``privately'' means that no third party
should be able to infer the identities of the parties taking part in
the communication (i.e. $A$ and $B$) (PAP Goal 1).
Moreover, if $A$ wants to communicate with $B$ but $A\not\in S_B$,
 the protocol should also conceal $B$'s identity (and presence) to
$A$ (PAP Goal 2).
A run of the protocol in which $A$ wants to communicate with $B$
proceeds as follows:
\begin{enumerate}
\item $A$ generates a nonce $N_A$. Then, $A$ prepares a message
  $M=\{``hello\text{''},N_A,K_A \}_{K_B}$, and broadcasts
  $(``hello\text{''},M)$.
\item When a principal $C$ receives message $(``hello\text{''},M)$, it performs
  the following three steps:
\begin{enumerate}
\item
  $C$ tries to decrypt $M$ with its own private key. If the
  decryption fails, (which implies that $C\neq B$), then $C$ creates a
  ``decoy'' message $\{N\}_K$ (creating a random $K$, and keeping
  $K^{-1}$ secret), broadcasts $(``ack\text{''},\{N\}_K)$ and finish
  its execution. If decryption succeeds, then $C=B$ (and so from now
  on we will refer to $C$ as $B$). $B$ then continues to the next
  step.
\item

$B$ checks that $A\in S_B$. If this fails, i.e. $A\not\in S_B$,
  then $B$ creates a ``decoy'' message $\{N\}_K$, broadcasts
  $(``ack\text{''},\{N\}_K)$ and finishes execution. Otherwise $B$
  continues to the next step.

\item
Finally, $B$ generates a fresh nonce $N_B$, and broadcasts the message\\
  $(``ack\text{''},\{``ack\text{''}, N_A,N_B,K_B\}_{K_A})$.
\end{enumerate}

\end{enumerate}
It is interesting to see the use of ``decoy'' messages, to prevent
attacks in which an intruder $I$ prepares a message
$M=\{``hello\text{''},N_C,K_A \}_{K_B}$, impersonating principal $A$. If
decoy messages were not present, then $I$ would send
$(``hello\text{''},M)$, and deduce whether $A\in S_B$ by noticing a
response from $B$. However, using decoys only helps to confuse an
attacker doing traffic analysis, and breaks down when considering a
``timed'' intruder, as we will show in the next section.

\subsubsection*{An Attack Over an Implementation of the Private Authentication protocol}

We show an attack in which $I$ can find whether $A\in S_B$. The attack
is illustrated in Figure \ref{attack}, where $I$ is trying to attack
$A$, $B$ and $C$, which since $I$ does not know their identities are
called $X$, $Y$ and $Z$. First, suppose that $I\not\in S_B$ (the
attack for the case in which $I\in S_B$ is analogous). First $I$ needs
to know how long, on average, it takes to $B$ to compute each step of
the protocol as described above.  To discover this $I$ could prepare
various messages:

\begin{figure}
\begin{center}
\input{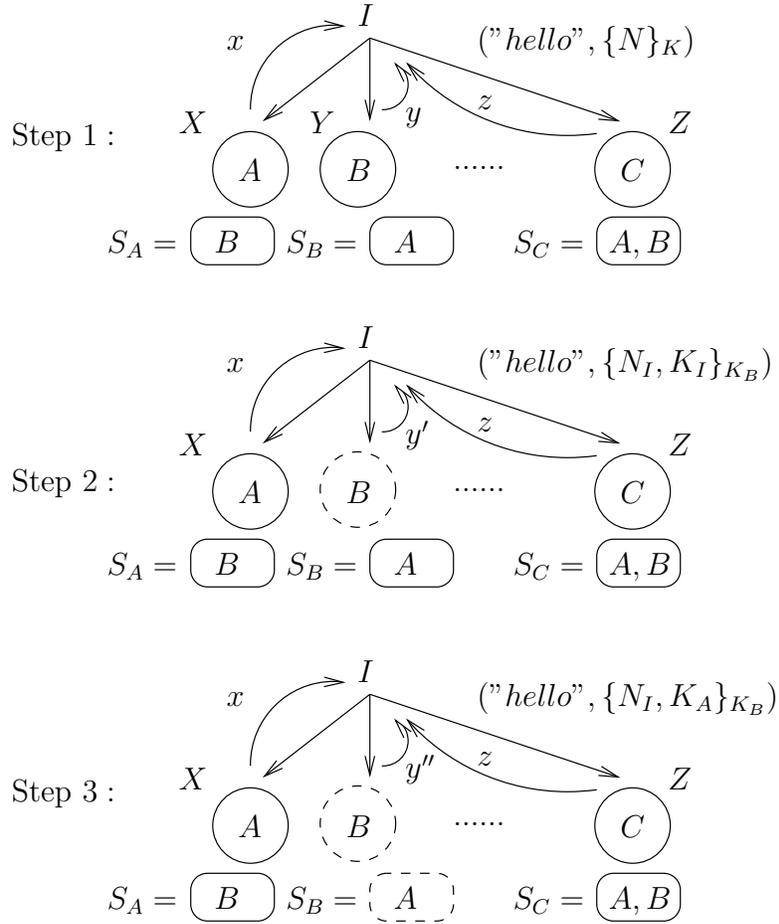}
\caption{The attack over the PAP protocol, where $X$, $Y$ and $Z$ are unknown identities by the intruder $I$. $A$, $B$ and $C$ are real identities, with corresponding $S_A$, $S_B$ and $S_C$ sets; $x$, $y$, $y'$, $y''$ and $z$ are timing values (Dashed circles indicate the intruder's knowledge of inner values)}
\label{attack}
 \end{center}
\end{figure}

\begin{enumerate}
\item

Firstly, $I$ sends a message $(``hello\text{''},\{N\}_K)$, where $K$
  is not the public key of any other participant. This would generate
  a number of decoy responses from the other participants, which $I$
  can time (Step 1 in Figure \ref{attack}, for times $x$, $y$ and
  $z$).

\item
 Secondly, $I$ sends a message
  $(``hello\text{''},\{``hello\text{''},N_I, K_I\}_{K_B})$. Again,
  this generates decoy responses from the other parties which $I$ can
  time (Step 2 in the figure).  However, if $B$ is present,
  then \emph{one} response will have longer time (ie. we get times
  $x$, $z$ and $y'$ with $y'$ longer than $y$), reflecting the
  successful decryption and check that $I\not\in S_B$ performed by $B$
  (Recall we assume that $I\not\in S_B$). Up to this point, $I$ has
  information that allows him to infer $B$'s presence (hence the
  dashed circle in the figure); Thus, this attack already violates
  goal 2 of Abadi's requirements \cite{Aba00}: $B$ should protect its
  presence if a party $X$ is willing to communicate with $B$ but
  $X\not\in S_B$ (Step 2).
\item

 Finally, if $B$ is present, then $I$ sends message
  $(``hello\text{''},\{``hello\text{''},N_I, K_A\}_{K_B})$. This would
  generate again the same decoy responses, except one that takes
  longer (Step 3 in the figure, for $x$, $z$ and $y''$ with $y''$
  longer than $y'$). If this response takes the same time as the above
  item, then $I$ can deduce that $A\not\in S_B$. Otherwise, if the
  response takes longer (reflecting the nonce generation $N_B$ and
  encryption performed by $B$) then $I$ can deduce that $A\in S_B$.
\end{enumerate} 

If $I\in S_B$, then the second step above returns the \emph{longest}
time, and the third message would take either \emph{less} time or
equal.

After recording this information, $I$ has three time values $t_0, t_1$
and $t_2$. $t_0$ corresponds to the time in which $B$ is not present;
$t_1$ corresponds to the time in which $B$ is present but its
communicating party $X \not\in S_B$. Finally, $t_2$ corresponds to the
case in which $B$ is present and its communicating party $X\in S_B$.
With these values at hand, now an attacker can check if $A\in S_B$ for
an arbitrary $A$.

Timing in networks is often accurate but if the accuracy is too low,
the intruder can repeat the timings (i), (ii) and (iii) and perform
statistical analysis to increase the probability of the inferences to
be correct~\cite{Koc96}. We propose this as future work, when we have a
probabilistic, timed model checker at disposal.

\section{Related Work}
\label{sec:related-work}
Many approaches focus on the study of
protocols that use timestamps~\cite{DG04,ES00,Lowe97,BP98,GLM03}.
Recent work of Delzanno et.al. \cite{DG04} presents an automatic
procedure to verify protocols that use timestamps, like the Wide
Mouthed Frog protocol. In their work, differently from ours, a global
clock is assumed, and timeouts and retransmissions are not discussed.
Evans and Schneider \cite{ES00} present a framework for timed
analysis.  Differently from our (UPPAAL) model checking, it is based
on a semi-decision procedure with discrete time.  In that work, the
usage of retransmissions is hinted at as future work, but not (yet)
addressed.  Lowe \cite{Lowe97} also analyses protocols with timing
information; his work shares with us the model checking approach,
although Lowe's approach is based on a discrete time model. A global
clock is also assumed, and timeouts nor retransmissions are addressed.
Closer to ours is the work of Gorrieri et al.~\cite{GLM03}, in which a
real-time process algebra is presented for the analysis of
time-dependent properties. They focus on compositionality results, and
no model checking is presented. Gorrieri et al. also show how timeouts can
be modelled, although retransmissions are not discussed.


Regarding our 
timing attack upon Abadi's protocol, Focardi et al~\cite{FGLM02}
develop formal models for Felten and Schneider's \emph{web privacy}
timing attack~\cite{Fel00}; their modelling activity shares with our
work the idea of using timed automata for analysis, although our
attack illustrates a timing attack over a ``pure'' security protocol.


\section{Conclusions}
Security protocol analysis is a mature field of research that has
produced numerous results and valuable tools for the correct
engineering of security protocols. Despite a large body of literature
on the subject, most analysis methods do not take time into
consideration (with the exception of a few papers, considering mainly
the use of timestamps). We argue that this is not realistic as all
distributed protocols need to implement timeouts (possibly followed by
retransmissions).

In this paper we address some of the issues that need to be
considered when including time-related parameters in the engineering
process of a security protocol.

Our first contribution is a method for the design and analysis of
security protocols that consider timing issues. We model security
protocols using timed automata, and use UPPAAL to simulate, debug and
verify security protocols in a real time setting. To this end, we
employ a general Dolev~Yao style intruder (naturally encoded in
UPPAAL), and we remark that modelling the intruder as a timed automata
implicitly extends its power to take into account the time
sensitivity. Our method allows us to specify security protocols in
detail, with timeouts and retransmissions. This increases the
confidence in the analysis, since the modelled protocols are closer to
their implementations than the classical analysis
(e.g. CASPER~\cite{Lowe97} or the constraint-based methods of
\cite{MS01,CE02}).

Secondly, by analyzing the protocols in the Clark and Jacob library
and the SPORE library, we see that most protocols schemas (w.r.t.\
timeouts) fit into a small number of common patterns.  We analyse the
efficiency and security of each of the patterns. Still, as possible
future work we would like to perform a full UPPAAL analysis of each of
these literature protocols (just as we do for Yahalom in
Section~\ref{sec:real-protocol}).

Other novel and more real-life protocols which are sensitive to timing
issues (e.g. besides timeouts, use for instance puzzles) may benefit
from our analysis, e.g. the Host-Identity-Protocol (HIP), initially
analysed by Aura et al.~\cite{ANG05}.

Our third contribution is an illustration of the implicit information
carried by timing. The mere act of sending a message at a specific
moment in time, and not another, carries information. We propose a
novel security protocol that exploits this fact to achieve
authentication. The protocol replaces the standard nonces with timed
challenges, which must be replied at specific moments in time to be
successful. Although it is a preliminary idea, it exposes clearly the
fact that security protocols can use and take advantage of time.

Finally we address threats specifically involving timing should also
be considered; specifically, timing attacks. We illustrate these
attacks in the context of security protocols, where branching allows
an intruder to deduce information that is intended to be kept
secret. Specifically, we mount an attack over an imperfect
implementation of Abadi's private authentication protocol
\cite{Aba00}. Solutions to avoid timing attacks in the implementations 
are usually expensive (e.g. noise injection or branch
equalization), and it is not our purpose to investigate them. Here we
merely lift the known problem of timing attacks, typically mounted
against the cryptosystem to obtain secrets keys, to security protocols in
general where the information leakage can be, in principle, anything.

One possible direction of future work is to consider a timed and
probabilistic model checker (in the lines of \cite{Arg01}
or \cite{KNP02b}), that would allow us to study the protocols of
Section 6. Moreover, a probabilistic setting would allow us to model,
more realistically, the network latency. This, in turn, would provide
us with a finer method to tune sensitive timing values. Another
possible direction for future research would be to implement a
compiler from a meta notation (similar to the standard notation, plus
timing information) supporting symbolic terms, to UPPAAL
automata. Ultimately, these directions of future work would contribute
to a method of secure systems engineering.

\paragraph{Acknowledgements}
Gabriele Lenzini, Conrado Daws, Jeroen Doumen, Jerry den Hartog, Ari
Saptawijaya and the anonymous reviewers of FMSE2004 provided useful
comments.

\bibliography{refs,protocol,timing} \bibliographystyle{plain}


\end{document}